\documentclass[twocolumn,aps,prl,floatfix,superscriptaddress]{revtex4-1}

\usepackage{hyperref}
\usepackage{amsmath}
\usepackage{amsfonts}
\usepackage{amsbsy}
\usepackage{xcolor}
\usepackage{soul}
\usepackage{graphicx}

\newcommand{\ket}[1]{\ensuremath{\left\vert #1 \right\rangle}}

\newcommand{\spvec}[1]{\ensuremath{\mathbf{#1}}}

\newcommand{\unitvec}[1]{\hat{\mathbf{{#1}}}}

\newcommand{\cgshort}[3]{\ensuremath{\mathcal{C}_{#1,#2}^{(#3)}}}

\newcommand{\EQREF}[1]{Eq.~(\ref{#1})}
\newcommand{\bea}{\begin{eqnarray}}
\newcommand{\eea}{\end{eqnarray}}
\newcommand{\beq}{\begin{equation}}
\newcommand{\eeq}{\end{equation}}

\newcommand{\Rv}{{\bf R}}

\newcommand{\bE}{{\bf E}}

\newcommand{\pola}{\hat{\mathbf{e}}}
\newcommand{\rv}{{\bf r}}
\newcommand{\br}{{\bf r}}

\newcommand{\Dc}{{\cal D}}

\newcommand{\eo}{\epsilon_0}
\newcommand{\<}{\langle}
\renewcommand{\>}{\rangle}
\renewcommand{\(}{\left(}
\renewcommand{\)}{\right)}
\renewcommand{\[}{\left[}
\renewcommand{\]}{\right]}

\newcommand{\Pc}{\mathcal{P}}
\newcommand{\Pcv}{\boldsymbol{{\cal P}}}

\newcommand{\commentout}[1]{{}}

\newcommand{\cbE}{\boldsymbol{\mathbf{\cal E}}}

\usepackage{float}

\begin{document}
\title{Storing light with subradiant correlations in arrays of atoms}
\date{\today}
\author{G. Facchinetti}
\affiliation{Mathematical Sciences, University of Southampton, Southampton SO17 1BJ, United Kingdom}
\affiliation{\'Ecole Normale Sup\'erieure de Cachan, 61 avenue du Pr\'esident Wilson, 94235 Cachan, France}
\author{S. D. Jenkins}
\affiliation{Mathematical Sciences, University of Southampton, Southampton SO17 1BJ, United Kingdom}
\author{J. Ruostekoski}
\affiliation{Mathematical Sciences, University of Southampton, Southampton SO17 1BJ, United Kingdom}

\begin{abstract}
We show how strong light-mediated resonant dipole-dipole interactions between atoms can be utilized in a control and storage of light.
The method is based on a high-fidelity preparation of a collective atomic excitation  in a single correlated subradiant eigenmode in a lattice.
We demonstrate how a simple phenomenological model captures the qualitative features of the dynamics and sharp transmission resonances that may find applications in sensing.
\end{abstract}

\maketitle

Resonant emitters play a key role in optical devices for classical and quantum technologies.
Atoms have particular advantages because of an excellent isolation from environmental noise with well-specified resonance frequencies and no absorption due to nonradiative losses.
At high densities, however, they exhibit strong light-mediated resonant dipole-dipole (DD) interactions that can lead to uncontrolled and unwanted phenomena, such as resonance broadening, shifts and dephasing. According to common wisdom, these are considered as a design limitation in quantum and classical light technologies, e.g., in  quantum metrology~\cite{Nicholson_clock,Ye2016}, sensing~\cite{BudkerRamalisNatPhys2007}, information processing~\cite{HAM10},  in the storage of light and in the implementations of quantum memories~\cite{LiuEtAlNature2001,FleischhauerLukinPRL2000,ChanelierePhotStorNAT2005,ChoiEtAlNature2008}.
DD interactions also receive significant attention, e.g., in Rydberg gases~\cite{TongEtAlPRL2004,HeidemannEtAlPRL2007,PritchardEtALPRL2010,WilkEtAlPRL2010,SchemppEtAlPRL2010}.
Here we show how strong radiative interactions can be harnessed in engineering long-living \emph{collective} excitations that open up avenues for utilizing resonant DD interactions in the control and storage of light, and in sensing. Our protocol is based on controlled preparation of large, many-atom subradiant excitations, where the light-mediated interactions between the atoms strongly suppress radiative losses.

Superradiance~\cite{Dicke1954} where the emission of light is coherently enhanced in an ensemble of emitters has continued to attract considerable interest~\cite{GrossHarochePhysRep1982} with the recent experiments focusing
on light in confined geometries~\cite{kimblesuper}, weak excitation regime~\cite{Roof16,Araujo16,wilkowski}, and the related shifts of the resonance frequencies~\cite{Keaveney2012,Meir13,ROH10,Jenkins_thermshift,Jennewein_trans}. Its counterpart, subradiance, describes coherently suppressed emission
due to a weak coupling to the radiative vacuum. Because of the weak coupling, subradiant states are challenging to excite and have experimentally proved elusive. In atomic and molecular systems subradiance has been observed in pairs of trapped ions~\cite{DeVoe} and molecules~\cite{Hettich},  as well as in weakly bound ultracold molecular states~\cite{McGuyer,Takasu}. In a large atom cloud a subradiant decay was recently observed in the long tails of a radiative decay distribution~\cite{Guerin_subr16} that indicated a small fraction of the atoms exhibiting a suppressed emission.

In our model, an incident light excites a collective atomic state that exhibits a significant radiative vacuum coupling. The excitation is then transferred to a radiatively
isolated cooperative state. The cold atoms that store the light excitation are confined in a planar lattice, providing a protection against nonradiative losses--that typically are a common hindrance to observation of subradiance. The state transfer is achieved by rotating the collective atomic polarization by an effective magnetic field. Depending on the size of the lattice and the confinement of the atoms, we find substantially suppressed radiative emission where up to 98-99\% of the total excitation is transferred into a \emph{single} subradiant eigenmode of the interacting multiatom system. The correlated many-atom excitation spatially extends over the entire lattice and is therefore fundamentally different from two-atom subradiant states~\cite{DeVoe,McGuyer,Takasu}. We develop a simple phenomenological two-mode model that provides an intuitive description of the light storage dynamics, and qualitatively captures the essential features, e.g., of the Fano resonance of the forward-scattered light.

\begin{figure}
	\centering
		\includegraphics[width=0.95\columnwidth]{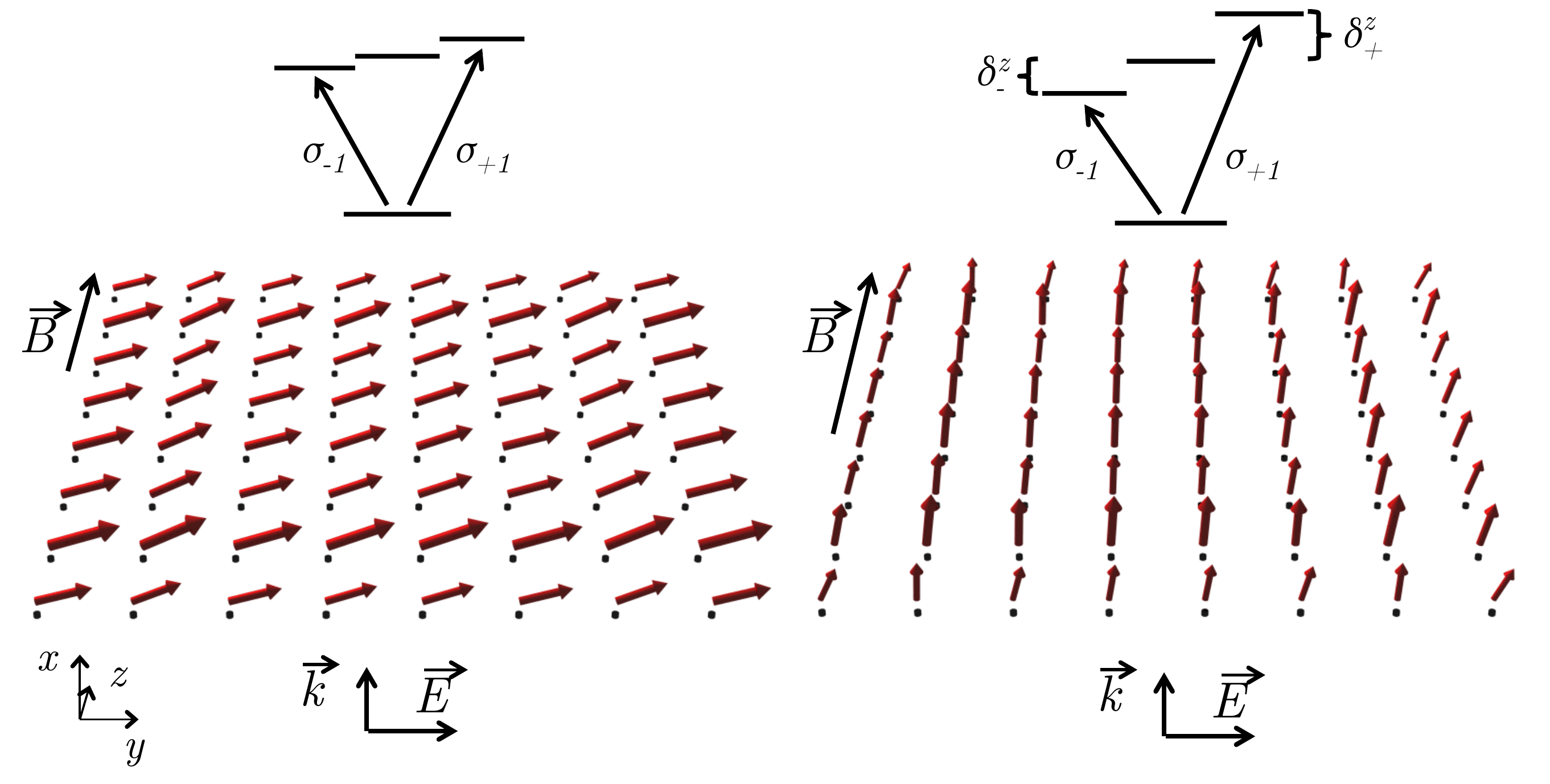}
    \vspace*{-11pt}
		\caption{
		Schematic illustration and numerically calculated
		response. The atoms (one per site) are confined in a small square 2D  $8\times 8$ array on the $yz$ plane. The linearly polarized
		(along $y$) incident light propagates along the positive $x$ direction driving the $|J=0,m_J=0\>
		\rightarrow |J'=1,m_J=\pm1\>$ transitions. The arrows represent the numerically calculated steady-state atomic dipoles at each site.
		A real or synthetic magnetic field along the $z$ axis induces Zeeman shifts,
		effectively rotating the dipoles around the $z$ axis. For $(\delta^{z}_{+},\delta^{z}_{-}) = (0.1,0.3)\gamma$ (left panel) this rotation is small, but for
		$(\delta^{z}_{+},\delta^{z}_{-}) = (0.45,1.75)\gamma$ (driven at the resonance of the subradiant mode; right panel), the dipoles are
		oriented approximately normal to the lattice, representing a collective excitation eigenmode with a factor of 50 narrowed linewidth.
 }
    \vspace*{-11pt}
		\label{figure_lattice}
\end{figure}

We consider a tightly-confined square planar array of atoms (e.g., a 2D optical lattice) with one atom per site (Fig.~\ref{figure_lattice}).
The light-induced radiative DD interactions lead to collective behavior of the
atoms in the lattice that is dramatically different from the response of an individual, isolated atom~\cite{Jenkins2012a,Castin13,Bettles_lattice,Yoo16,Rey_Weyl}.
The atoms are either at fixed positions or we address the position fluctuations using the model of a finite optical lattice with the potential depth $sE_R$ in the units of the lattice photon recoil energy $E_R$~\cite{Morsch06,SOM}.
In the numerics, the lattice spacing $a=0.55\lambda$, except when specified otherwise.
Whenever we consider a finite lattice depth, we take the confinement normal to the lattice $\simeq0.12 a$.
The atoms are illuminated by an incident weak-intensity laser with the amplitude
$\cbE(\rv)= {\cal E}_0(y,z) \pola_y \exp(ikx)$, with polarization $\pola_y$ and ${\cal E}_0(y,z)$ either constant or
a Gaussian profile on the $yz$ plane.
Here, and in the rest of the paper, all the field amplitudes and the atomic polarization correspond to the slowly varying positive
frequency components with oscillations at the laser frequency $\omega$.
We consider a near-resonance $J=0\rightarrow J'=1$ atomic transition (e.g.\ Yb, Sr) and assume a controllable Zeeman level
splitting of the $J'=1$ manifold. The Zeeman shifts could be induced
by magnetic fields or, e.g., by AC Stark shifts~\cite{gerbier_pra_2006}.

In the numerical simulations we calculate the optical response by evaluating all the multiple scattering
events~\cite{Ruostekoski1997a,Morice1995a} between the atoms in an array. In the limit of low light intensity, for stationary atoms the results
are exact~\cite{Javanainen1999a,Lee16}, and we also include the vacuum fluctuations of the atomic positions in the lowest vibrational
level of each lattice site~\cite{Jenkins2012a}. This is done by stochastically sampling the atomic positions at each site in each realization according to the density
distribution and then ensemble-averaging the results.
At each stochastic run we have the $N$ atoms fixed at positions $\rv_j$, and
we calculate the dipole moment $ {\bf d}_j= \Dc \sum_{\sigma} \pola_{\sigma} \Pc_{\sigma}^{(j)}$ for each atom $j$, where $\Dc$ denotes the reduced dipole matrix element.
Each atom has three polarization amplitude components $\Pc_{\sigma}^{(j)}$ associated with the unit circular polarization vectors
$\pola_{\pm1}=\mp (\pola_x\pm i \pola_y)/\sqrt{2}$ and $\pola_0=\pola_z$, that are coupled with the transitions
$|J=0,m=0\>\rightarrow |J'=1,m=\sigma\>$.

In the limit of low light intensity, the excited state
population of the atoms vanishes and the excitation amplitudes satisfy~\cite{Jenkins2012a,Lee16}
\beq
  \label{eq:dynamics_pol}
 \frac{d}{dt} \Pc^{(j)}_\sigma  = \left( i \Delta_\sigma - \gamma \right)
  \Pc^{(j)}_\sigma + i\frac{\xi}{{\cal D}} \unitvec{e}_\sigma^{\ast} \cdot
  \eo \spvec{E}_{\rm ext}(\rv_j) \textrm{,}
\eeq
where  $\xi=6\pi\gamma/k^3$ and the single-atom
Wigner-Weisskopf linewidth $\gamma=\Dc^2 k^3/(6\pi\hbar\epsilon_0) $.
The detuning from the atomic resonance
$\Delta_{\sigma} = \omega-\omega_{\sigma}=\omega-(\omega_0+ \sigma\delta^{z}_\sigma)$ where $\omega_0$ is the resonance frequency of
the $|J=0\>\leftrightarrow |J'=1,m=0\>$ transition and $\pm\delta^{z}_{\pm}$ are the shifts of the $m=\pm1$ levels (Fig.~\ref{figure_lattice}).
Each amplitude in Eq.~\eqref{eq:dynamics_pol} is driven
by the sum of the incident field and the fields scattered from all the
other $N-1$ atoms $\bE_{\rm ext}(\br_j) = \cbE (\br_j)+\sum_{l\neq j}
\bE^{(l)}_S(\br_j)$. The scattered dipole radiation field from the atom
$l$ is $\epsilon_0\bE^{(l)}_S(\br)={\sf G}({\bf r}-{\bf r}_l) \Dc \sum_{\sigma} \pola_{\sigma} \Pc_{\sigma}^{(l)}$, where
${\sf G}$ is the dipole radiation kernel, such that $\bE^{(l)}_S(\br)$
represents the electric field at $\br$ from
a dipole $\Dc \sum_{\sigma} \pola_{\sigma} \Pc_{\sigma}^{(l)}$  residing at $\br_l$~\cite{Jackson}.

We first consider a single, isolated atom. This is obtained in Eq.~\eqref{eq:dynamics_pol}  by setting
$ \spvec{E}_{\rm ext}(\rv_j)\rightarrow \cbE (\br_j)$.  The $y$-polarized light then drives the atomic
polarization components $ \Pc_{\pm 1}^{(j)}$ (Fig.~\ref{figure_lattice}). Here we instead write the equations of motion
in the Cartesian basis $ {\bf d}_j/\Dc = \pola_{x} \Pc_{x}^{(j)} +\pola_{y} \Pc_{y}^{(j)} +\pola_{z} \Pc_{z}^{(j)}$, such that
$ \Pc_{x}^{(j)}=( \Pc_{-1}^{(j)}- \Pc_{+1}^{(j)})/\sqrt{2}$ and $ \Pc_{y}^{(j)}=-i( \Pc_{-1}^{(j)}+ \Pc_{+1}^{(j)})/\sqrt{2}$.
We obtain
\begin{align}
\dot\Pc^{(j)}_x & = (i \Delta_0-i\tilde\delta -\gamma) \Pc^{(j)}_x - \bar\delta  \Pc^{(j)}_y,\\
\dot \Pc^{(j)}_y & = (i \Delta_0-i\tilde\delta -\gamma) \Pc^{(j)}_y + \bar\delta  \Pc^{(j)}_x +i\xi\eo {\cal E}_0/\Dc\,,
\label{eq:singleatom}
\end{align}
where $\tilde\delta =(\delta^{z}_{+}-\delta^{z}_{-})/2$, $\bar\delta =(\delta^{z}_{+}+\delta^{z}_{-})/2$, and $\Delta_0$ denotes the detuning
of the $m=0$ state. The incident light directly drives only
 $\Pc^{(j)}_y $, but the energy splitting of the levels $|m=\pm1\>$ introduces a coupling between  $\Pc^{(j)}_x $ and  $\Pc^{(j)}_y $. Although
the incident field is perpendicular to $\Pc^{(j)}_x $, the light can therefore still excite $\Pc^{(j)}_x $ by first driving $\Pc^{(j)}_y $. The $J=0\rightarrow J'=1$
 transition is isotropic when the excited-state energies are degenerate and any orientation of the orthogonal basis also forms an eigenbasis.
For $\bar\delta\neq0$,  $\Pc^{(j)}_{x/y}$ no longer are eigenstates. The dipoles are consequently turned toward the $x$ axis by the rotation around the effective magnetic field.

\begin{figure}
	\centering
		\includegraphics[width=\columnwidth]{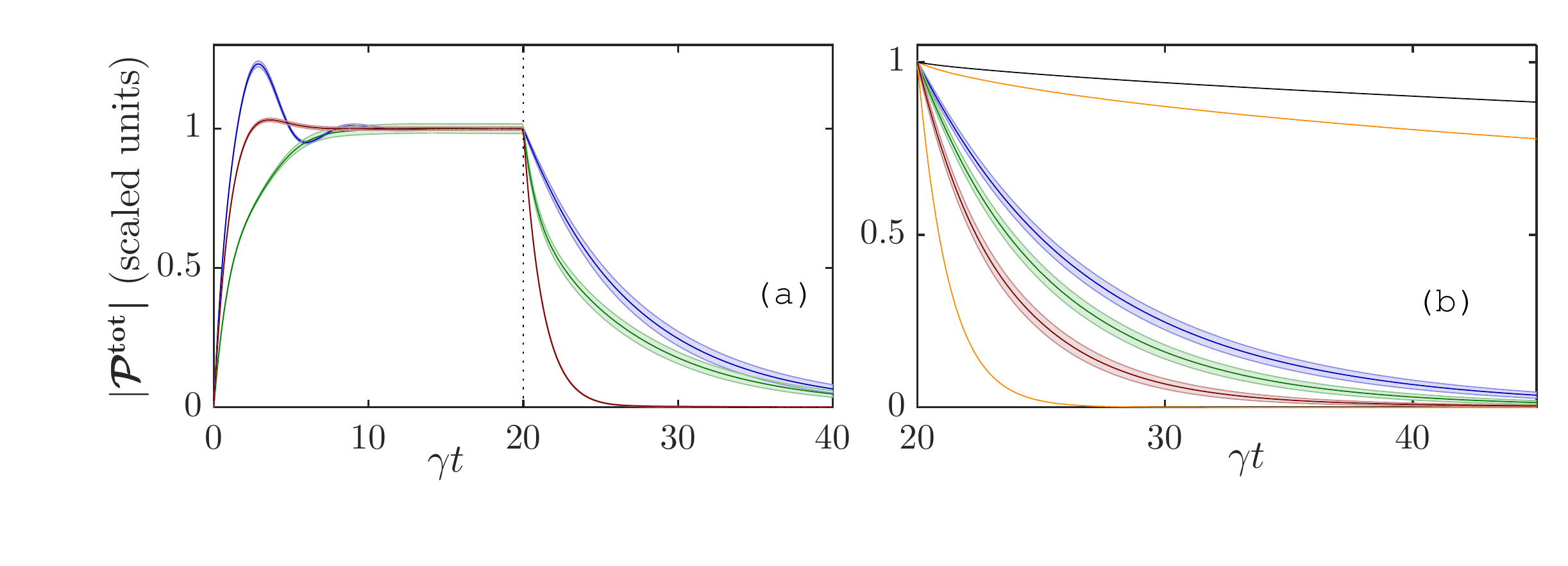}
    \vspace*{-30pt}
		\caption{ The dynamics of the total atomic polarization density in the lattice of 20$\times$20 sites for different Zeeman shifts and lattice heights. (a) Curves from slow to fast decay:
		 $({\delta}^{z}_{+},{\delta}^{z}_{-},\Delta_0) = (0,0,0)$, (0.4,0.6,0)$\gamma$, and (1.1,1.1,0.65)$\gamma$ ($s=50$). At $\gamma t=20$ (when each curve is normalized to one), the Zeeman shifts
		and the incident light are switched off. When the dipoles are oriented close to the $x$ axis
		(see Fig.~\ref{figure_lattice} on right),
		the decay is slow (collective subradiance). For ${\delta}^{z}_{\pm}=0$ the dipoles are pointing along the $y$ axis and decay rapidly.
		An exponential fitting provides decay rates $0.79\gamma$, $0.16\gamma$, and $0.14\gamma$.
		 (b) The curves from top: incident Gaussian beam with fixed atomic positions, plane-wave excitation for fixed atomic positions, for lattice with $s=50$, 20, 5, [$(\delta^{z}_{+},\delta^{z}_{-},\Delta_0) = (1.1,1.1,0.65)\gamma$],
		 and for fixed atomic positions
		 ($\delta^{z}_{\pm}=\Delta_0=0$).
		 The shaded lines around the curves represent the stochastic uncertainties of the excitation amplitudes due to the vacuum fluctuations of
		the atomic positions. }
		\label{fig:timedynamics}
\end{figure}

For the entire interacting many-body system
we numerically calculate the optical response
for different Zeeman shifts and lattice heights and show in Fig.~\ref{fig:timedynamics}(a) the dynamics
of the total polarization of the system $ |\Pcv^{\rm tot}| = | \sum_{j,k} \Pc^{(j)}_k \pola_k|/N$ [in all the numerical results, the polarization amplitudes are expressed
in the dimensionless form $\Pc\rightarrow \Dc \Pc k^3/(6\pi \eo {\cal E}_0)$].
The incident light excites the $y$ components of the atomic dipoles.
Analogously to the single atom case, the Zeeman shifts turn the polarization density toward the $x$ direction. At the resonance~\cite{SOM} $(\delta^{z}_{+},\delta^{z}_{-},\Delta_0) = (1.1,1.1,0.65)\gamma$ we find the dipoles almost entirely
along the $x$ direction \footnote{We found that the many-atom dynamics is practically identical when we utilize the symmetry of Eqs.~\eqref{eq:singleatom} and~\eqref{coll2modey} and
transform to any new set of parameters $\Delta_0\rightarrow\Delta_0 +\Delta$, ${\delta^z_+}\rightarrow\delta^z_++\Delta$, ${\delta^z_-}\rightarrow\delta^z_- -\Delta$, for some $\Delta$.}.
After the evolution has reached the steady state, the Zeeman shifts and the incident laser are
turned off, resulting in a decay of the excitations. We fit the exponential functions to the decay profiles to obtain
numerical estimates for the collective radiative linewidths that we later compare with the collective eigenvalues.
For  $\delta^{z}_{\pm}=0$, the dipoles are in the lattice plane and the radiative decay rate $0.79\gamma$ is close to the single atom linewidth.
However, for $(\delta^{z}_{+},\delta^{z}_{-},\Delta_0) = (1.1,1.1,0.65)\gamma$ we find strongly suppressed decay of $0.14\gamma$, indicating that the \emph{entire} collective radiative excitation is
dominated by subradiance.
This is very different from the observation of long tails of radiative decay where only an extremely small fraction of the total excitation exhibits enhanced lifetime~\cite{Guerin_subr16}.

The lattice confinement affects
the subradiant decay [Fig.~\ref{fig:timedynamics}(b)] and for more strongly fluctuating atomic positions we obtain faster decay rates with $0.18\gamma$ and $0.28\gamma$ for $s=20$ and $s=5$,
respectively. For the case of fixed atomic positions a better fit is obtained by a double exponential (reflecting the occupation of eigenmodes with different linewidths, as explained later)
$b_1 e^{-c_1t}+b_2 e^{-c_2t}$, with $b_1\simeq 0.72$, $c_1\simeq 0.0032\gamma$, $b_2\simeq 0.24$, $c_2\simeq 0.027\gamma$.
The decay is dominated by an exponent that is about 300 times smaller than the one for a single atom.

Owing to the resonant DD interactions the atoms respond collectively to light, exhibiting collective
excitation eigenmodes with distinct collective radiative linewidths and line shifts. We can qualitatively understand the response by
analyzing the behavior of the most dominant modes. The incident light is phase-matched to
a smoothly-varying, phase-coherent excitation of the atoms. The linear polarization couples to a collective (``coherent in-plane'') mode in
which all the dipoles are coherently oscillating along the $y$ direction with the excitation $ \Pc_I$
--a collective eigenmode of the system in the absence of the Zeeman shifts. Since all the dipoles in this mode are in the lattice plane,
$ \Pc_I$ is responsible for strong reflection and transmission of light. For nonzero Zeeman shifts the mode no longer is an eigenmode, but as in the single atom case, the polarization
of the atoms is then turned toward the $x$ axis. This reorientation can be qualitatively analyzed by a simple two-mode model when we assume that
$ \Pc_I$ is predominantly coupled with a phase-coherent collective (``coherent perpendicular'') excitation $ \Pc_P$ where all
the atomic dipoles are oscillating in phase, normal to the plane~\cite{SOM}.  Also this mode is a collective eigenmode for $\delta^{z}_{\pm}=0$.
We can now establish an effective two-mode dynamics~\cite{SOM}
\begin{subequations}
\begin{align}
\dot\Pc_P & = (i \Delta_P-i\tilde\delta -\upsilon_P) \Pc_P - \bar\delta  \Pc_I, \label{coll2modex}\\
\dot \Pc_I & = (i \Delta_I-i\tilde\delta -\upsilon_I) \Pc_I +  \bar\delta  \Pc_P +i\xi\eo {\cal E}_0/\Dc\,,
\label{coll2modey}
\end{align} \label{bothtwo}
\end{subequations}
where $\upsilon_{P/I}$ are the collective linewidths of the corresponding eigenmodes of the many-atom system (for $\delta^{z}_{\pm}=0$) and $\Delta_{P/I}=\omega-\omega_{P/I}=\Delta_0+\delta_{P/I}$ are the detunings of the incident light from the resonances of these modes (that are shifted by $\delta_{P/I}$).

The excitation $ \Pc_P$ dominantly radiates within the plane, enhancing interactions between the atoms: For light to escape, it generally undergoes many scattering events, so that the collective mode becomes strongly subradiant. After the excitation is driven into $ \Pc_P$, we set ${\cal E}_0=\bar\delta=0$, and the decay becomes slow. The light can be released by applying a fast
$\pi$-Rabi-pulse using $\bar\delta\neq 0$ that transfers the excitation back to $ \Pc_I$.

We also calculate the eigenmodes when $\delta^{z}_{\pm}=0$ for the full interacting system of atoms and light, and analyze the occupations of the different eigenmodes in
the steady-state responses of Fig.~\ref{fig:timedynamics}(b) (at $\gamma t=20$).
We use the occupation measure $L_j=| \mathrm{v}_j^T \mathrm{b}|^2/\sum_i | \mathrm{v}_i^T  \mathrm{b} |^2$
for the eigenvector $\mathrm{v}_j$ in the state $ \mathrm{b}$.
The resonance linewidths are then compared with the calculated decay
rates of Fig.~\ref{fig:timedynamics}(b).
We find that the steady-state excitation of the $\delta^{z}_{\pm}=0$ fixed atomic position case is dominated by the collective $ \Pc_I$ excitation eigenmode with about 50\% of the total excitation~\cite{SOM}. Its linewidth $\upsilon_I\simeq 0.79\gamma$ almost perfectly matches with the
fitted decay rate $0.80\gamma$ in Fig.~\ref{fig:timedynamics}(b). For the $(\delta^{z}_{+},\delta^{z}_{-},\Delta_0) = (1.1,1.1,0.65)\gamma$ fixed atomic position case the fitting
of the radiative decay to a double-exponential in Fig.~\ref{fig:timedynamics}(b) provided  a much better result.
This slowly-decaying case is dominated by the subradiant $ \Pc_P$ excitation eigenmode with about 70\% of the total excitation~\cite{SOM}.
The linewidth $\upsilon_P \simeq 3.1\times 10^{-3}\gamma$ indicates a strongly subradiant excitation and again very closely matches with the dominant exponent
$3.2\times 10^{-3}\gamma$ of the decay in Fig.~\ref{fig:timedynamics}(b). The reason for the double-exponential decay in Fig.~\ref{fig:timedynamics}(b) is a prominent
excitation $\sim15\%$ of an additional eigenmode whose linewidth $\simeq0.015\gamma$ notably differs from that of $ \Pc_P$.

Although the subradiant eigenmode with $\upsilon_P \simeq 3.1\times 10^{-3}\gamma$ has a uniform phase profile, its amplitude is smaller close to the lattice edges~\cite{SOM}.
This suggests that even a more targeted excitation of this mode can be achieved using a focused Gaussian laser beam. Indeed, a Gaussian beam with the standard deviation $6a$ increases
the occupation to 98\% of the total excitation~\cite{SOM}. The corresponding dynamics provides an excellent fit to a \emph{single} exponential with a decay rate of $3.1\times 10^{-3}\gamma$.

\begin{figure}
	\centering
    \vspace*{11pt}
		\includegraphics[width=0.95\columnwidth]{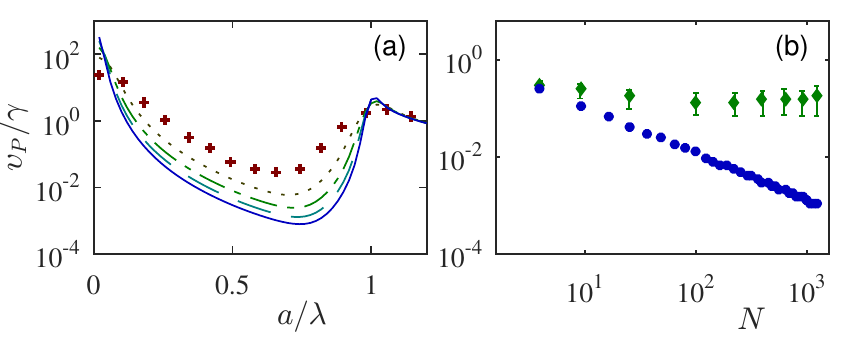}
    \vspace*{-10pt}
		\caption{The resonance linewidth $\upsilon_{P}$ of the subradiant collective eigenmode where the atomic dipoles coherently point to the normal of the lattice. (a) The dependence on the lattice spacing $a$. The curves from top: arrays
5$\times$5, 10$\times$10, 15$\times$15, 20$\times$20, 25$\times$25. (b) The dependence on the number of atoms for $a = 0.55\lambda$: $s=50$ (top curve), the atoms at fixed positions (lower curve).  For instance, $\upsilon_P \simeq 1.0\times 10^{-3}\gamma$ for the 35$\times$35 lattice in the lower curve. In comparison, the most subradiant eigenmode in this case has a linewidth $1.5\times 10^{-4}\gamma$.}
		\label{fig:size}
\end{figure}

The many-body nature of the light-mediated interactions manifests itself in a strong dependence of the suppressed decay on the size of the system.
In Fig.~\ref{fig:size}(b) we show the linewidth $\upsilon_{P}$ as a function of the atom number $N$.
For fixed atomic positions the mode becomes increasingly more subradiant in larger lattices with $\upsilon_P/\gamma \simeq N^{-0.91}$.
The fluctuations of the atomic positions suppress the linewidth narrowing and, e.g., $s=50$
has the large array limit $\upsilon_P \simeq 0.15\gamma$. Using tight confinement  in the Lamb-Dicke regime $\ell_j\ll a$, e.g., by optical tweezers, can significantly increase the lifetime of the subradiant state in large systems.

By varying the lattice spacing for different atom numbers we find that $\upsilon_{P}$ has a minimum around
$a/\lambda$=0.7-0.8 [Fig.~\ref{fig:size}(a)]. Around the minimum $\upsilon_{P}$ is also the most subradiant linewidth of the system.
The engineered excitations have particularly narrow linewidths for far red-detuned optical lattices for which $a/\lambda\agt 0.55$.

\begin{figure}
	\centering
    \vspace*{11pt}
		\includegraphics[width=0.95\columnwidth]{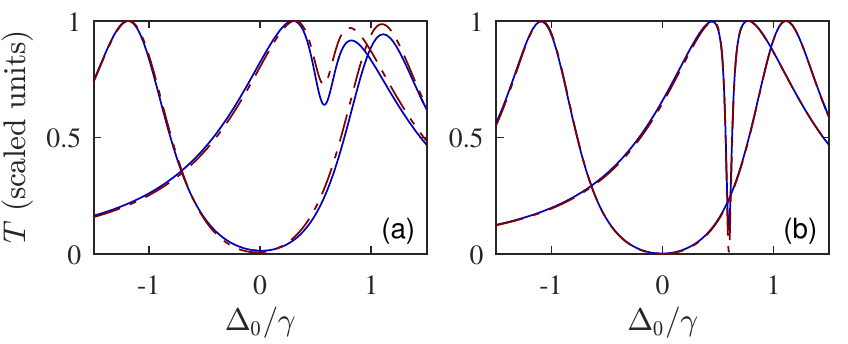}
    \vspace*{-10pt}
		\caption{The spectrum of forward or back scattered light and Fano resonances for different orientations of the dipoles:
		$({\delta}^{z}_{+},{\delta}^{z}_{-}) = (0.1, 0.2)\gamma$ (the orientation not far from the lattice plane; narrow resonances), $(0.45, 1.75)\gamma$
		(the orientation approximately normal to the plane; broad resonances) for a lattice of fixed atomic positions and (a) 3$\times$3; (b) 20$\times$20 sites.
		Numerical simulation (blue, solid curves), two-mode model (red, dashed curves) with numerically calculated eigenvalues for the two dominant eigenmodes $(\delta_P +i\upsilon_P)/\gamma\simeq
-0.65+0.0031i$ (20$\times$20), $-0.62+0.12i$ (3$\times$3),  and $(\delta_I+i\upsilon_I)/\gamma\simeq
-0.68+0.79i$ (20$\times$20), $-0.59+0.83i$ (3$\times$3).
The scattering resonance is approximately at the effective resonance of the subradiant mode [Eq.~\eqref{coll2modex}] $\Delta_P-\tilde\delta\simeq0$. }
		\label{fig:spectrum}
\end{figure}

The narrow linewidth $\upsilon_{P}$ manifests itself also in the resonance of the scattered light (Fig.~\ref{fig:spectrum}).
We display the spectrum of the steady-state response of the forward scattered light into a narrow cone of $|\sin\theta|\alt 0.1$. The full numerical simulation is compared with the two-mode model of Eqs.~\eqref{bothtwo} that qualitatively captures the main features of the spectra, indicating that the resonance behavior is dominated by the two collective modes.
The spectra exhibit a Fano resonance due to a destructive interference between different scattering paths
that involve either the excitation  $ \Pc_I$ only, or a scattering via $ \Pc_P$, as in $ \Pc_I \rightarrow \Pc_P\rightarrow \Pc_I$.
One can see from Eqs.~\eqref{bothtwo}~\cite{SOM} that the forward or back scattered light is suppressed when $\bar\delta^2\gg \upsilon_P\upsilon_I$ and that the resonances correspond to high (low) occupations of
$\Pc_P$ ($\Pc_I$) excitations.
In the limit that $\Pc_P$ is not strongly driven, the narrow spectral resonance is a direct
consequence of its subradiant linewidth in a large lattice (the resonances strongly depend on the lattice size; Fig.~\ref{fig:spectrum}), and the interference is analogous to the interference of bright and dark modes in
the electromagnetically-induced transparency (EIT)~\cite{FleischhauerEtAlRMP2005}. If $\Pc_P$ is strongly excited by the Zeeman shifts, the resonance notably broadens
and its width can be approximated by $[(\upsilon_I^2+4\bar\delta^2)^{1/2}-\upsilon_I]/2$ (for $\upsilon_P/\bar\delta\simeq0$)~\cite{SOM}.
In the limit of a large lattice the optical response varies between a full transmission ($ \Pc_P$ resonance) and complete reflection ($\bar\delta=0$)~\cite{SOM}.
Narrow transmission resonances due to collective radiative interference may also
be achieved in magnetodielectric solid-state resonator systems~\cite{CAIT}, and EIT in an optical lattice has been proposed~\cite{Bettles_lattice}.

In conclusion, we showed that collective light-atom interactions can be harnessed for a controlled preparation of a single, spatially-extended, multiatom subradiant excitation eigenmode, storing the incident light.
The possibility to engineer optical interactions may be promising, e.g., for the control of many-atom light shifts in lattice clocks~\cite{Nicholson_clock,Ye2016}, and our subradiant state exhibits suppressed shifts~\cite{SOM}. Moreover, the narrow resonance features are very sensitive to the Zeeman shifts and could also provide a detection mechanism of weak magnetic fields~\cite{SOM}. Unlike in a magnetometry using EIT~\cite{Fleisch_magneto} in weakly interacting vapors, the width of the resonance here is not limited by the single atom linewidth, but by the much narrower collective subradiant linewidth, resulting, e.g., in a sharp dispersion at the transmission resonance (Fig.~\ref{fig:spectrum}) and a large group delay.

\begin{acknowledgments}
We acknowledge financial support from the EPSRC
and the use of the IRIDIS High Performance Computing Facility at the University of Southampton.
\end{acknowledgments}

\appendix

\setcounter{equation}{0}
\setcounter{figure}{0}
\renewcommand{\theequation}{A\arabic{equation}}
\renewcommand{\thefigure}{A\arabic{figure}}

\section{Supplemental material}

\subsection*{Array of atoms}

We assume the atoms to be in a planar square array. We label the sites of the  $N_y\times N_z$ lattice with the lattice spacing $a$ as $i=1,\ldots, N$, $N=N_y N_z$.
The array could be formed by optical tweezers, optical lattice potential, or by nanofabrication. In the case of an optical lattice we approximate
the lattice site Wannier function $\phi_i(\rv) \equiv \phi(\rv - \Rv_i)$, centered at ${\bf R}_i$, by the ground-state
wavefunction of a harmonic oscillator, with the vibrational frequency $\omega = 2\sqrt{s}E_R/\hbar$ and the $1/e$ radius $\ell=\ell_y=\ell_z= as^{-1/4} / \pi$
of the atom density $\rho_i(\rv) \equiv |\phi_i(\rv)|^2$, where $E_R = \pi^2\hbar^2/(2ma^2)$ is the lattice-photon recoil energy and each site has a potential depth
$sE_R$~\cite{Morsch06}. In the $yz$ plane the atoms become more localized as the lattice depth $s$ is increased.
We assume that the confinement of the atoms around the $x=0$ plane results in $1/e$ radius of the atom density equal to $\ell_x$. In this work, whenever we consider
fluctuating atomic positions, we take $\ell_x\simeq 0.12 a$.

We assume one atom per site. In an optical lattice precise occupation numbers can be achieved by using ultracold atoms in a Mott-insulator state~\cite{Morsch06}.
Experimentally, the atoms in the Mott state can be prepared to a single-atom site-occupancy
even in situations where the lattice is superposed with a harmonic trapping potential by
manipulating the sites with excess occupancy~\cite{singlespin}.

\subsection*{Calculation of eigenmodes}

The collective radiative excitation eigenmodes of the full system of $N$ atoms can be solved writing the coupled system of atoms and light [Eq.~(1) in the main text]
as  $\dot{\mathrm{b}} = i \mathcal{H}\mathrm{b} + \mathrm{F}$, where $\mathrm{b}$ is a vector made of the amplitudes $\Pc^{(j)}_k$ and
$\mathrm{F}$ represents the external driving of the dipoles by the incident light~\cite{JenkinsLongPRB}. The coupling matrix  $i\mathcal{H}$
provides the light-induced interactions between the atoms and the first terms on the right-hand-side of Eq.~(1) in the main text.
The matrix $\mathcal{H}$ has $3N$ eigenmodes $\mathrm{v}_j$ with the eigenvalues $\delta_j + i \upsilon_j$
where $\delta_j=\omega_0-\omega_j$ is the shift of the collective mode resonance $\omega_j$
from the single atom resonance
and $\upsilon_j$ is the collective radiative linewidth.

\begin{figure}
	\centering
		\includegraphics[width=0.95\columnwidth]{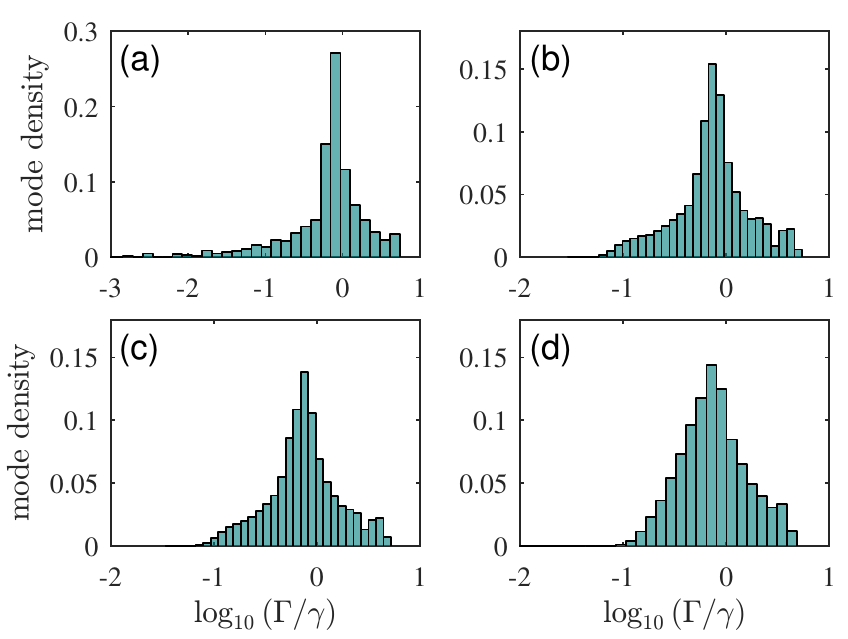}
    \vspace*{-10pt}
		\caption{Probability distributions (normalized to one) of collective radiative eigenmode resonance linewidths  for (a) fixed atomic
		positions, (b)  the optical lattice height $s=50$, (c) $s=20$,  (d) $s=5$.
		 The distributions become narrower as the atoms are less tightly confined. This is particularly manifested by the disappearance of the long tails of strongly subradiant
		 eigenmode decay rates  in the distribution as the lattice becomes shallower.
		}
    \vspace*{-11pt}
		\label{fig:eighist}
\end{figure}

\begin{figure}
	\centering
		\includegraphics[width=0.95\columnwidth]{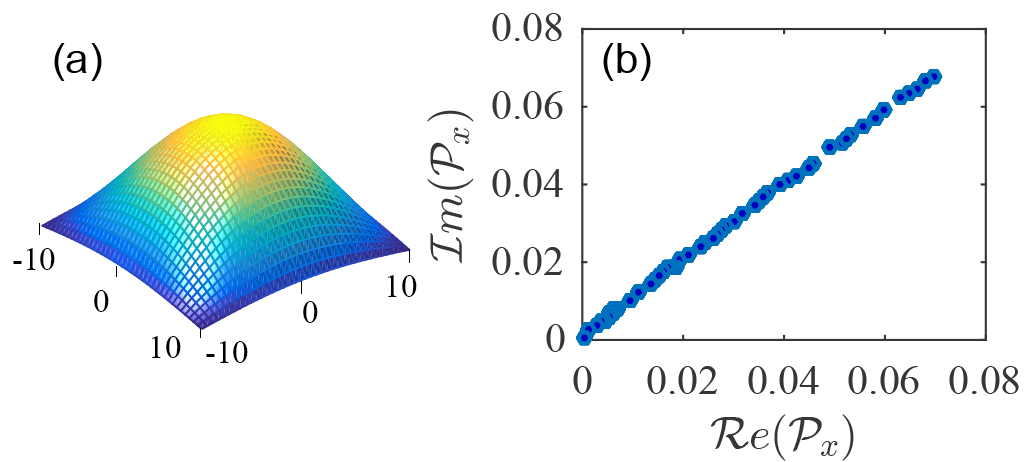}
    \vspace*{-10pt}
		\caption{The collective subradiant eigenmode where the atomic dipoles are coherently oscillating in the $x$ direction for the case of fixed atomic positions. (a) The absolute value of the amplitude of the $x$-component of the atomic polarization density (arbitrary linear scale) on the array plane. The excitation is largest at the center of the lattice and decreases toward the edges. The corresponding $y$ and $z$ components are approximately zero.
		The array plane for the eigenfunction amplitude is shown in the units of the lattice sites.
		(b) The $x$-component of the complex polarization density for each lattice component that exhibits high phase coherence over the entire lattice, indicated by the narrow distribution of the phase values for the atoms. }
    \vspace*{-11pt}
		\label{fig:submode}
\end{figure}

In Fig.~\ref{fig:eighist} we show the probability distributions of the collective eigenmode resonance linewidths. The increase in the size of the lattice results in a larger number of
subradiant eigenmodes with notably narrower linewidths. The collective subradiant eigenmode where the atomic dipoles are coherently oscillating in the $x$ direction is displayed in
Fig.~\ref{fig:submode}; the distribution of the complex amplitudes demonstrates the phase coherence over the entire lattice.

Although the eigenvectors here form a basis, they are generally not orthogonal, since $\mathcal{H}$ is not Hermitian. For the isotropic $J=0\rightarrow J'=1$ system $\mathcal{H}$ is
symmetric, and we can determine the biorthogonality condition $\mathrm{v}_j^T \mathrm{v}_i=\delta_{ji}$, except for possible zero-binorm states for
which $\mathrm{v}_j^T \mathrm{v}_j=0$ (that we have not encountered in our system).
We therefore define an overlap measure $L_j=| \mathrm{v}_j^T \mathrm{b}|^2/\sum_i | \mathrm{v}_i^T  \mathrm{b} |^2$
for the eigenvector $\mathrm{v}_j$ in the state $ \mathrm{b}$. We also use this measure to determine the eigenmodes that are the closest to the ideal phase-coherent
polarization excitations $ \Pc_I$ and $ \Pc_P$.

\begin{figure*}
	\centering
		\includegraphics[width=1.5\columnwidth]{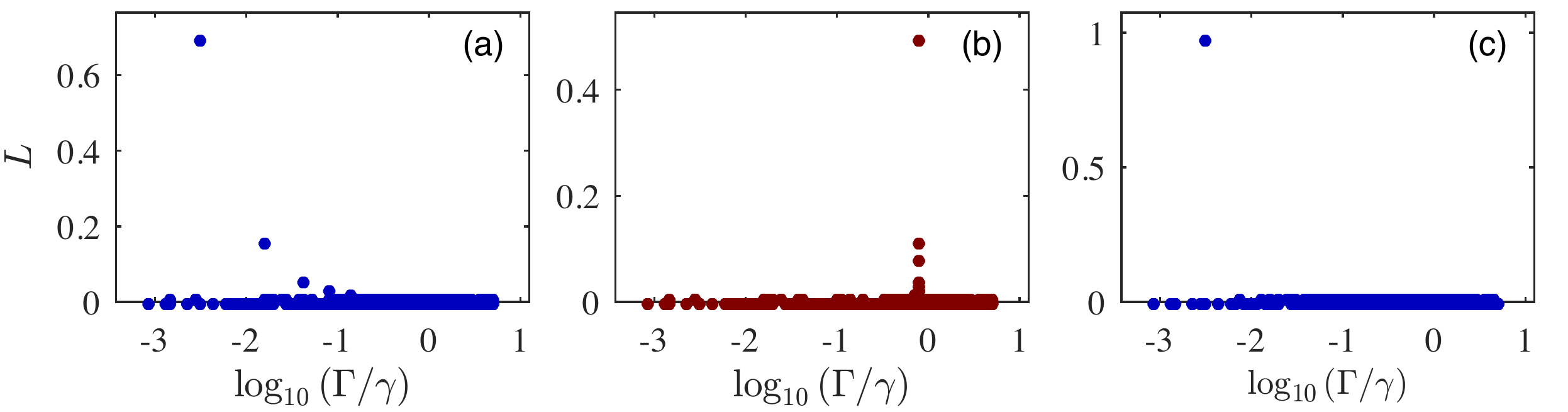}
    \vspace*{-8pt}
		\caption{The measure of the mode population $L$ of the different eigenmodes in the steady-state response for fixed atomic positions,
		(a) incident plane wave excitation, $(\delta^z_{+}, \delta^z_{-},\Delta_0) = (1.1,1.1,0.65)\gamma$; (b) incident plane wave excitation, $(\delta^z_{+}, \delta^z_{-},\Delta_0) = (0,0,0)\gamma$; (c) Gaussian beam excitation with the standard deviation $6a$, $(\delta^z_{+}, \delta^z_{-},\Delta_0) = (1.1,1.1,0.65)\gamma$. The eigenmodes are ordered by their collective radiative resonance linewidths on a logarithmic scale.
		In (b) the atomic polarization density is in the lattice plane and the excitation is dominated
		(over 50\% of the total excitation) by the collective mode where the dipoles are coherently oscillating in the $y$ direction. In (a) and (c) the atomic polarization density is pointing normal to the lattice plane,
		and the excitation is dominated by about (a) 70\%, (c) 98\% by the collective subradiant mode where the atomic dipoles are coherently oscillating in the $x$ direction. In (a) also some modes with notably
		broader resonances are occupied that manifests themselves in a faster initial decay of the radiative excitation (see Fig.~2 in the main text).
    }
    \vspace*{-11pt}
		\label{fig:popu}
\end{figure*}
In Fig.~\ref{fig:popu} we show the populations of the different eigenmodes in the steady-state response for the case where the subradiant mode with the polarization
vectors normal to the lattice plane is excited and the case where it is not. In all the cases only a small number of modes are significantly excited. In the case of  driving of the dipole excitation normal to the lattice using a plane wave the excited eigenmodes have different linewidths resulting in an approximate double-exponential decay rate that is described in the main text (see Fig.~2 in the main text). For the Gaussian beam driving, only one eigenmode is notably excited. This is because the Gaussian beam intensity is better matched with the density distribution of the eigenmode in Fig.~\ref{fig:submode}.

\begin{figure}
	\centering
		\includegraphics[width=0.9\columnwidth]{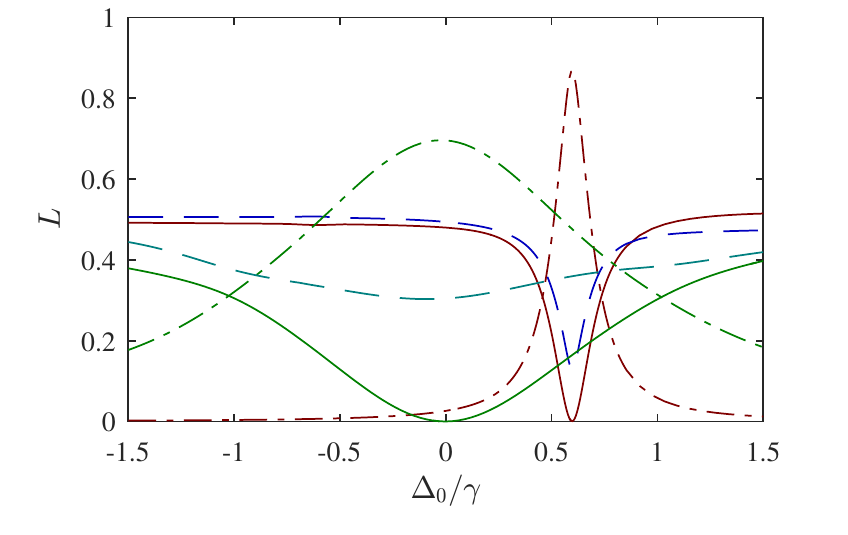}
    \vspace*{-12pt}
		\caption{The measure of the eigenmode populations $L$ in the steady-state response  as a function of the frequency of the incident light for fixed atomic positions. The occupations correspond to the spectrum of the forward or back scattered light shown in Fig.~4(b) of the main section.
		 $(\delta_{+}^z, \delta_{-}^z) = (0.1, 0.2)\gamma$, sharp resonances: the collective subradiant eigenmode where the atomic dipoles are coherently oscillating in the $x$ direction
		 (red dash-dotted line), the collective eigenmode where the atomic dipoles are coherently oscillating in the $y$ direction (red solid line), the sum of the mode populations
		 of all the other 1198 eigenmodes (blue dashed line). $(\delta_{+}^z, \delta_{-}^z) = (0.45, 1.75)\gamma$, broad resonances: the same curves with green colors.
		}
    \vspace*{-11pt}
		\label{fig:popspe}
\end{figure}

In Fig.~\ref{fig:popspe} we display the eigenmode populations of the main eigenmodes in the steady-state response as a function of the incident light frequency. These populations correspond to the scattered light spectra shown in Fig.~4(b) of the main text. The peak of the subradiant mode excitation represents the case where the light scattering in the forward or back direction vanishes. In the limit of an infinitely large lattice with a subwavelength lattice spacing and fixed atomic positions only the exact forward or back scattering is possible, since in that case only the zeroth order Bragg peak survives.

\begin{figure}
	\centering
		\includegraphics[width=0.78\columnwidth]{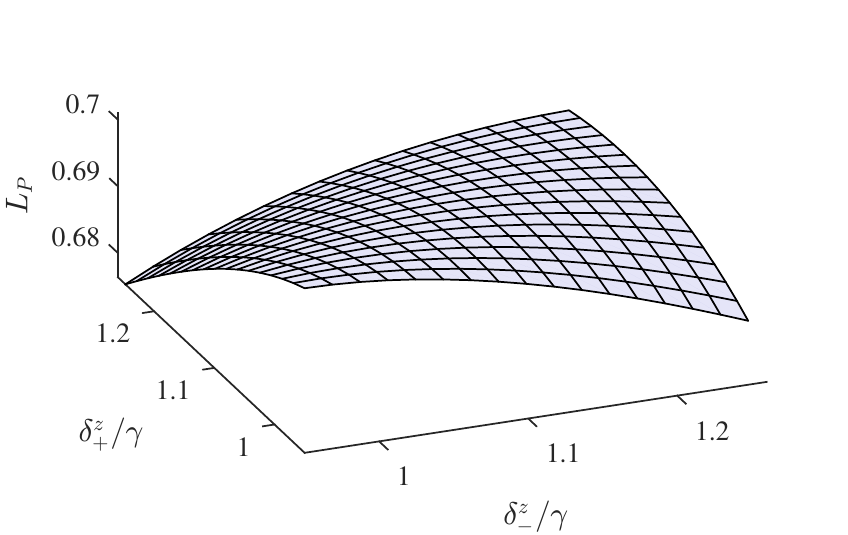}
    \vspace*{-10pt}
		\caption{
		The measure of the population $L_P$ of the coherent subradiant eigenmode where the dipoles are pointing in the direction normal to the lattice  in the steady-state response to a plane-wave excitation  as a function of the Zeeman shifts; $\Delta_0 = 0.65$, with the fixed atomic positions.  The excitation $\Pc_P$ is maximized around $\delta^z_{+}\simeq\delta^z_{-}$,
    $\Delta_0 =  -\delta_P$. }
    \vspace*{-11pt}
		\label{fig:planezeeman}
\end{figure}

\begin{figure}
	\centering
		\includegraphics[width=0.78\columnwidth]{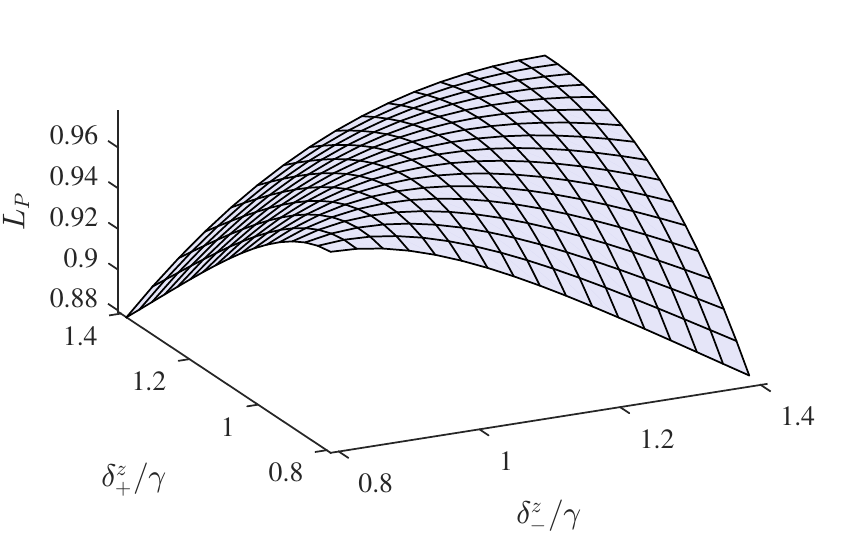}
    \vspace*{-10pt}
		\caption{
		The same as Fig.~\ref{fig:planezeeman}, but for a Gaussian beam excitation. }
    \vspace*{-11pt}
		\label{fig:gausszeeman}
\end{figure}

The populations of the subradiant eigenmode excitation $\Pc_P$ as a function of the Zeeman shifts are shown  for both plane-wave (Fig.~\ref{fig:planezeeman}) and Gaussian (Fig.~\ref{fig:gausszeeman})  incident fields.
The maximum in the figure can be found using the effective two-mode model (see below). The full numerical calculation reveals high populations of the mode even for small values of $\delta_{\pm}^z$, as e.g., shown in the peak of the curve in Fig.~\ref{fig:popspe}. Around $(\delta_{+}^z, \delta_{-}^z) = (0.05, 0.1)\gamma$ we find close to 95\% excitation of the subradiant mode even for a plane-wave incident field and about 99\% excitation for the Gaussian beam.
Small values of $\delta_{\pm}^z$ require notably longer evolution times before reaching the steady state. At small $\delta_{\pm}^z$, the large occupation values are also more sensitive to
the precise field profile due to the edge effects of the lattice.

\subsection*{Two-mode model}

\begin{figure}
	\centering
		\includegraphics[width=0.8\columnwidth]{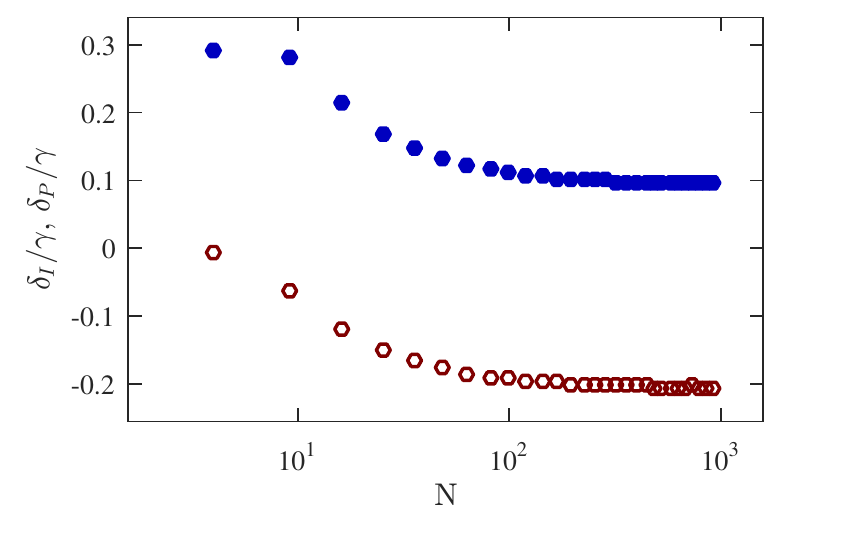}
    \vspace*{-14pt}
		\caption{
		The resonance shifts $\delta_P$ (blue solid circles) and $\delta_I$ (red rings) of the coherent perpendicular and in-plane collective modes, respectively, as a function of the atom number
		for fixed atomic positions with the lattice spacing $a=0.735\lambda$. For the strongly subradiant collective mode with the dipoles coherently pointing normal to the plane, the resonance
		shifts become increasingly suppressed as the atom number increases.}
		\label{fig:shift}
\end{figure}

In the main section we introduced a phenomenological two-mode model in which case the two polarization density amplitudes $\Pc_P$ and $\Pc_I$ obey the simplified dynamics given by
 \begin{subequations}
  \begin{align}
    \dot\Pc_P &= (i(\Delta_0+\delta_P-\tilde{\delta})-\upsilon_P)\Pc_P - \bar{\delta}\Pc_I, \label{eq:Pmode} \\
   	\dot\Pc_I &= (i(\Delta_0+\delta_I-\tilde{\delta})-\upsilon_I)\Pc_I + \bar{\delta}\Pc_P + f \,,
	\label{eq:Imode}
  \end{align}
  \label{eq:twomode}
\end{subequations}
where $\tilde\delta =(\delta^{z}_{+}-\delta^{z}_{-})/2$, $\bar\delta =(\delta^{z}_{+}+\delta^{z}_{-})/2$ are defined in terms of the Zeeman shifts, and $\Delta_0$ denotes the detuning
of the $m=0$ state.
The linear polarization of the incident field with the driving $f=i\xi\eo {\cal E}_0/\Dc$ [$\xi=6\pi\gamma/k^3$, with $\gamma=\Dc^2 k^3/(6\pi\hbar\epsilon_0) $] couples to the ``coherent in-plane'' collective eigenmode with the amplitude $ \Pc_I$ (resonance shift $\delta_I$, linewidth $\upsilon_I$) in
which all the atoms are coherently excited along the $y$ direction. The ``coherent perpendicular'' eigenmode with the amplitude $ \Pc_P$ (resonance shift $\delta_P$, linewidth $\upsilon_P$)  represents  a mode
where all the atomic dipoles are oscillating in phase and pointing normal to the plane. Example behavior of the resonance shifts of the two collective eigenmodes as a function of the lattice size are given in Fig.~\ref{fig:shift}. In the limit of a large array the resonance shifts of the subradiant eigenmode become more suppressed. The control of many-atom light shifts in lattice clocks is of particular importance~\cite{Nicholson_clock,Ye2016}, as these can limit the performance of the clocks.
\begin{figure}
	\centering
		\includegraphics[width=0.97\columnwidth]{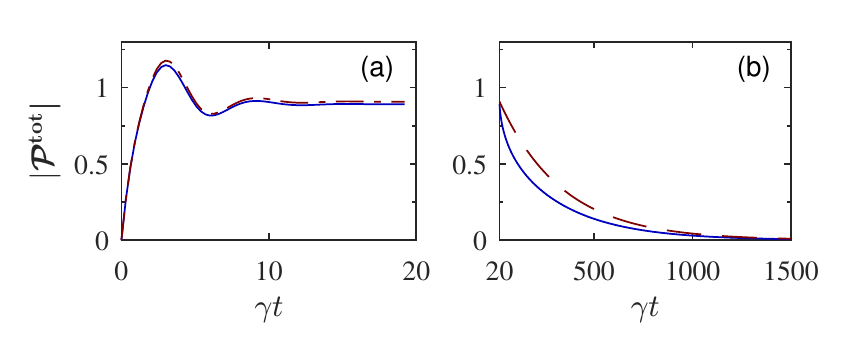}
		\includegraphics[width=0.97\columnwidth]{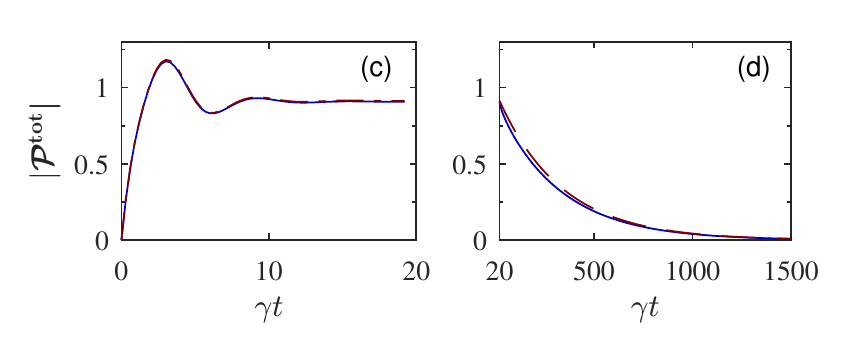}
    \vspace*{-12pt}
		\caption{Comparisons between the phenomenological two-mode model and the exact numerical simulation. The Zeeman shifts and the detuning $(\delta_{+}^{z}, \delta_{-}^{z},\Delta_0)=(1.1,1.1,0.65)\gamma$ have been chosen such that the collective polarization in the steady-state response is pointing normal to the lattice plane. In the two-mode model we have used the numerical values of the full eigenvalue calculation  that are $\delta_P = -0.65\gamma$, $\delta_I = -0.68\gamma$, $\upsilon_P = 0.0031\gamma$, $\upsilon_I=0.79\gamma$. The initial evolution of the laser-driven lattice before reaching the steady-state distribution for the case of (a) plane-wave excitation, (c) Gaussian beam excitation. The evolution after the incident light and the Zeeman shifts are turned off for the case of (b) plane-wave excitation, (d) Gaussian beam excitation. In the case of a plane-wave excitation the two-mode model
		differs from the full numerical solution at early times due to the contribution of additional collective modes in the dynamics (see Fig.~\ref{fig:popu}). The additional modes decay faster and the dynamics of the calculations is more similar at later times when  the slowly decaying subradiant coherent perpendicular mode dominates.
 }
		\label{fig:2modevsexact}
\end{figure}

The two-mode model qualitatively captures many of the essential features of the full many-body dynamics. This is illustrated in the spectra of Fig.~4 in the main text that
shows the Fano resonances of the forward (or back) scattered light. In Fig.~\ref{fig:2modevsexact} we also show the comparison between the dynamics given by the two-mode model and the full numerics of all the 1200 collective excitation eigenmodes. For the plane-wave excitation the decay rates of the two cases differ at early times owing to the notable contribution of collective eigenmodes with different decay rates in Fig.~\ref{fig:popu}(a). In the case of a Gaussian incident field excitation, the agreement is better, since in that case the entire excitation is dominated by a single collective eigenmode Fig.~\ref{fig:popu}(c).

We may also easily calculate the steady-state solution of Eqs.~\eqref{eq:twomode}
\begin{subequations}
\begin{align}
  \label{eq:ssI}
 	\Pc_I &= -i\frac{Z_P(\Delta_0)}{\bar{\delta}^{2} - Z_P(\Delta_0)Z_I(\Delta_0)}f \\
 	\Pc_P &= -i\frac{\bar{\delta}}{Z_P(\Delta_0)}\Pc_I \textrm{,}
\end{align}
\end{subequations}
where we define
\begin{equation}
  \label{eq:ZDef}
  Z_{P/I}(\Delta_0) \equiv \Delta_0 + \delta_{P/I} -\tilde{\delta} + i \upsilon_{P/I} \textrm{.}
\end{equation}
The ratio of the amplitudes,
\begin{equation}
	\left|\frac{\Pc_P}{\Pc_I}\right| = \left|\frac{\bar{\delta}}{Z_P(\Delta_0)}\right|\,,
\end{equation}
indicates when the subradiant excitation $\Pc_P$ becomes dominant.
At the resonance $\Delta_0 + \delta_{P} - \tilde{\delta} = 0$ we have $|\Pc_P/\Pc_I|=|\bar\delta|/\upsilon_P$,
\begin{equation}
	\left|\Pc_P\right| = \left|\frac{\bar{\delta}}{\bar{\delta}^{2}+\upsilon_I\upsilon_P}f \right|, \quad
		\left|\Pc_I\right| = \left|\frac{\upsilon_P}{\bar{\delta}^{2}+\upsilon_I\upsilon_P}f \right|\,,
\end{equation}
where we have assumed $|\delta_P - \delta_I |\ll \upsilon_I$
so that we can neglect any difference between the $\Pc_I$ and
$\Pc_P$ resonance frequencies. (We find that this holds approximately true for the lattice $a\simeq 0.55\lambda$.)

\subsubsection*{Forward and back scattered light}

Since the dipoles of the $ \Pc_P$ excitation point in the direction normal to the plane, only the coherent in-plane collective mode can emit in the exact forward or back directions.
Consequently, the forward and back scattered light amplitudes are in the steady-state response proportional to $\Pc_I$ amplitudes, given by Eq.~\eqref{eq:ssI}. We express the reflectance and the transmittance amplitudes in terms of the incident and the scattered field amplitudes $ \spvec{E}_I$ and $ \spvec{E}_S$, respectively
  \beq
  	  r = \frac{\unitvec{d}\cdot
      \spvec{E}_S(-\unitvec{e}_x)}{ \unitvec{d} \cdot \spvec{E}_I(\unitvec{e}_x)},\quad
          t = \frac{\unitvec{d}\cdot\left(\spvec{E}_I(\unitvec{e}_x) +
        \spvec{E}_S(\unitvec{e}_x)\right)}{ \unitvec{d} \cdot
      \spvec{E}_I(\unitvec{e}_x)} \textrm{.}
  \eeq
  We then obtain from Eq.~\eqref{eq:ssI}
  \begin{align}
    r &= \frac{r_0\upsilon_I(\upsilon_{P}-i(\Delta_P -\tilde{\delta}))}{\bar{\delta}^{2}-(\Delta_P -\tilde{\delta} + i \upsilon_{P})(\Delta_I -\tilde{\delta} + i \upsilon_{I})} \textrm{ ,} \label{eq:r_amp} \\
    t &= 1 + r  \label{eq:t_amp}\,,
\end{align}
where $r_0$ denotes the reflectance amplitude at the resonance of the coherent in-plane collective mode when the Zeeman shifts vanish $\bar{\delta}=0$, and $\Delta_{P/I}\equiv \Delta_0 + \delta_{P/I}$.

We can express the power reflectance $|r|^2$ in the limit $\upsilon_P/\bar{\delta} \simeq 0$
\begin{equation}
  \begin{split}
  \label{eq:no}
 |r|^2 \simeq  \frac{{(r_0\upsilon_I)}^{2}\[(\Delta_{P}-\tilde{\delta})^{2}+\upsilon_P^2\]}{|(\Delta_{P}-\tilde{\delta})^{4}-(\upsilon_I^2-2\bar{\delta}^{2})(\Delta_{P}-\tilde{\delta})^{2}+ \bar{\delta}^{4}|} \textrm{ ,}
  \end{split}
\end{equation}
where we have assumed $|\delta_P - \delta_I |\ll \upsilon_I$. In this limit we can then analytically calculate the half-width at half maximum for this resonance and obtain
\begin{equation}
	w \simeq \frac{1}{2}\(\sqrt{\upsilon_I^2 + 4\bar{\delta}^{2}} - \upsilon_I\)\,.
	\label{width}
\end{equation}
This simple expression qualitatively explains the observed behavior of the spectral resonances of Fig.~4 in the main section. For the stronger driving of the $\Pc_P$ excitation by the Zeeman shifts $\bar{\delta}$ the resonance is significantly broadened. In the limit of a weak driving, the resonance narrows and eventually only depends on the very narrow resonance linewidth $\upsilon_P$.

The scattering problem can be further simplified when we consider an infinite lattice on the $yz$ plane for fixed atomic positions. The 2D lattice behaves similarly to a 2D diffraction grating. For the subwavelength lattice spacing only the zeroth order Bragg peak of the scattered light survives. This corresponds to the exact forward and back scattered light. The energy conservation then states that
$|r|^2+|t|^2=1$. Combining this with Eqs.~\eqref{eq:r_amp} and~\eqref{eq:t_amp} yields $r_0=-1$, indicating that an incident field at the resonance of the coherent in-plane collective mode experiences a total reflection when the Zeeman shifts are zero.

We can use this result to simplify the reflectance amplitude formula. We find a local minimum of the transmittance on
$\Delta_0 + \delta_{P} - \tilde{\delta} = 0$
\begin{equation}
  \label{eq:R_delta_M}
  r(-\delta_P + \tilde{\delta}) \approx - \frac{\upsilon_I\upsilon_P}{\bar{\delta}^2 + \upsilon_I\upsilon_P}
\end{equation}
where we have again assumed $|\delta_{P} - \delta_{I} | \ll \upsilon_I$. When the Zeeman shifts satisfy $\bar{\delta}^{2} \gg \upsilon_P\upsilon_I$, reflectance on
$\Pc_P$ resonance is suppressed, and transmittance is enhanced. This is illustrated in Fig.~4 of the main section where the resonances appear in the 20$\times$20 lattice. For the 3$\times$ 3 lattice the reflectance is only suppressed when $({\delta}^{z}_{+},{\delta}^{z}_{-}) = (0.45, 1.75)\gamma$, but not in the case of $({\delta}^{z}_{+},{\delta}^{z}_{-}) = (0.1, 0.2)\gamma$ when $\bar{\delta}^{2} \sim \upsilon_P\upsilon_I$.
Remarkably, in the limit of a large array the response can therefore vary between a complete reflection and full transmission. The complete reflection can be achieved more generally for an infinite 2D array of dipoles with a subwavelength lattice spacing, as shown previously for both resonator~\cite{CAIT} and atomic~\cite{Bettles_prl16} cases. 

\subsubsection*{Sensitivity to weak magnetic fields}

We show how the excitation of the collective subradiant state and the resulting narrow transmission resonance could be used in the detection of weak magnetic fields. The basic idea is closely related to the EIT magnetometry. However, the EIT magnetometry is based on the independent atom scattering, so the frequency scales for the suppression of absorption and the sharp dispersion are determined by the \emph{single-atom} resonance linewidths. The key element in the cooperative transmission is the interplay between the \emph{collective} subradiant excitation $ \Pc_P$ and the collective strongly radiating $ \Pc_I$. In the large arrays, the corresponding resonance linewidths satisfy $\upsilon_P \ll \upsilon_I$. This can lead to very narrow transmission resonances, determined by $\upsilon_P$, where the ratio between the dispersion and absorption is proportional to $1/\upsilon_P$.

EIT has been proposed as a method for measuring weak magnetic fields~\cite{Fleisch_magneto,FleischhauerEtAlRMP2005}. In a typical EIT setup of a lambda-three-level atom, the transition amplitudes from two electronic ground states $|a\>$ and $|b\>$ to the same electronic excited state $|c\>$ destructively interfere, resulting in a spectral transparency window where the absorption is suppressed. This is a single-atom effect where the atoms are assumed to respond to light independently.  For a small two-photon detuning between the levels $|a\>$ and $|b\>$, the electric susceptibility of the medium can be approximated by $\chi=\chi'+i \chi''$ with
\beq
\chi'  \simeq -{24\pi\gamma_{bc}\rho\Delta\omega_{ab}\over k^3\Omega^2},\quad
\chi'' \simeq {12\pi \gamma_{bc}\rho \Gamma_{ab}\over k^3\Omega^2}\,,
\eeq
where the imaginary part $\chi''$ leads to absorption and the real part $\chi'$ to dispersion. Here $\rho$ denotes the density of the atoms, $\Delta\omega_{ab}$ the two-photon detuning between the levels $|a\>$ and $|b\>$,
$\Omega={\bf d}\cdot {\bf E}_p/\hbar$ the pump laser Rabi frequency driving the $|a\>\leftrightarrow |c\>$ transition, and $\gamma_{bc}$ the single-atom resonance linewidth of the probe $|b\>\leftrightarrow |c\>$ transition. The decay rate for the loss of coherence between the levels $|a\>$ and $|b\>$ is given by  $\Gamma_{ab}$ and it results, e.g., from atomic collisions and the fluctuations of the laser.
Since the two-photon detuning depends on the magnetic Zeeman shifts between the levels $\Delta\omega_{ab}\simeq \mu_B(m_a g_a-m_b g_b)/\hbar$ (here $\mu_B$ is Bohr's magneton, $g_j$ is the Land\'{e} g-factor and $m_j$ is the magnetic quantum number for the level $j$), the dispersion is proportional to the applied magnetic field, while the absorption at the same time is suppressed by the EIT.
The phase shift of the light propagating a distance $L$ inside the sample is obtained by
\beq
\Delta \phi= k(n-1)L\simeq  -{2 \gamma_{bc}\Delta\omega_{ab}\over \Omega^2} \rho\sigma_{\rm cr} L\,,
\eeq
where we have used $n\simeq 1+\chi'/2$ for the index of refraction $n$, and $\sigma_{\rm cr}= 3\lambda^2/(2\pi)$ denotes the resonance cross-section.

The EIT magnetometry is based on achieving a large phase shift by propagating the light beam through a sufficiently large atom cloud, even when the pump field intensity, proportional to $\Omega^2$, is large. Close to the EIT resonance an otherwise opaque medium becomes almost transparent due to the destructive interference between the different atomic transitions. This results in the suppressed attenuation of the beam (given by $\chi''$) and allows a long propagation distance inside the sample.
The suppression of the absorption at the EIT resonance is associated with a sharp variation of the dispersion curve.

The collective resonant dipole-dipole interactions between the atoms in our array result in a similar type of interference effect between the \emph{collective} eigenmodes $ \Pc_P$  and $ \Pc_I$.  This is illustrated by the
reflection and transmission amplitudes of Eqs.~\eqref{eq:r_amp} and~\eqref{eq:t_amp} that display the resonances of Fig.~4 of the main text. The corresponding occupations of the collective modes are shown in Fig.~\ref{fig:popspe} and the width of the resonances can be estimated by \EQREF{width}.

We have not performed a full 3D array calculation, but we can still calculate the effect by considering several stacked layers of 2D arrays when we add the effect of different layers using the mean-field theory of standard optics. This can provide a reasonable approximation if the layers are not too close to each other. In the limit of a large lattice the $ \Pc_P$ excitation does not contribute to the coherent light propagation (since only the zeroth order Bragg peak exists), and we write the effective electric susceptibility as $\chi \simeq \< P\> /(\epsilon_0 {\cal E}_0) \simeq\rho {\cal D} \<{\cal P}_I\>/(\epsilon_0 {\cal E}_0)$, where the atom density can be given in terms of the layer separation $\Delta z$ as $\rho\simeq 1/(a^2\Delta z)$.
We assume that $\bar\delta^2 \gg \upsilon_P \upsilon_I$ and that the effective detuning from the $ \Pc_P$ excitation resonance $|\Delta_P-\tilde\delta|\ll \bar\delta$. Note, however, that since in the limit of a large array $\upsilon_P\rightarrow0$ and can be very small, we can still simultaneously have a very small $\bar\delta \ll \upsilon_I$. This corresponds to an extremely narrow transmission resonance in which case the sample is transparent at the exact resonance with a sharp variation of the dispersion curve.

We again ignore the differences between the resonance frequencies of the two modes $\delta_P \simeq \delta_I$. From the two-mode model solutions we then find
\beq
\chi'  \simeq {6\pi\gamma\rho(\Delta_P-\tilde\delta)\over k^3\bar \delta^2},\quad
\chi'' \simeq {6\pi\gamma\rho \upsilon_P\over k^3\bar \delta^2}\,.
\eeq
The corresponding phase shift is given by
\beq
\Delta \phi =  {\gamma (\Delta_P-\tilde\delta)\over \bar \delta^2} \rho \sigma_{\rm cr} L\,.
\eeq
Here any changes in both $\bar\delta$ and $\tilde\delta$ are sensitive to the magnetic field variation. We can set the maximum allowed length of the sample to be defined by limiting the attenuation of the beam by absorption [given by $\exp(-\pi\chi'' L /\pi)$]  to be
$\pi\chi'' L_{\rm max}  /\pi \sim 1$. For $L_{\rm max}$ the phase shift $\Delta\phi\sim (\Delta_P-\tilde\delta)/\upsilon_P$ can still be very large due to extremely narrow subradiant mode resonance $\upsilon_P$.

Analogously, we can calculate the group delay $\tau_g$ of a resonant light pulse for the 2D atom array due to cooperative response. This is given by the derivative of the
phase of the transmission amplitude with respect to the frequency
\beq
\tau_g = \left.\frac{d}{d\omega} \arg
    t(\omega)\right|_{\Delta_P-\tilde\delta=0}  \simeq {\upsilon_I\over \bar\delta^2}\,,
\eeq
where we have taken the limit of a large array $r_0=-1$ and assumed $\bar\delta^2 \gg \upsilon_P \upsilon_I$ with $\delta_P \simeq \delta_I$. By using the value of $\upsilon_P\sim 10^{-3}\gamma$ for the $35\times35$  lattice, we can obtain a sharp resonance even for $\bar\delta^2\sim 10^{-2} \gamma^2$, and $\tau_g\sim 100/\gamma$ (since $\upsilon_I\sim \gamma$). The cooperative effect is even more dramatic for the $150\times 150$ array for which $\upsilon_P\sim 10^{-4}\gamma$ (obtained from $\upsilon_P/\gamma\sim N^{-0.91}\gamma$; see the main section), and
the group delay becomes significant $\tau_g\sim 10^3/\gamma$ even for a narrow 2D plane of atoms.

The analogy between the collective array response and the EIT could also be extended beyond magnetometry, e.g, to slow and stopped light propagation~\cite{LiuEtAlNature2001,DuttonHau,dutton_prl_2004}.

\subsubsection*{The effect of the atomic level structure}

The specific analysis of the subradiant excitation was obtained for the $J=0\rightarrow J'=1$ transition. Such level structure can commonly be found in cold-atom
experiments that use alkaline-earth-metal atoms or rare-earth metals, such as Yb or Sr. For alkaline-metal atoms the level structure is more complicated. For $^{87}$Rb the optical
response was experimentally and theoretically studied for the $F=2\rightarrow F=3$ transition~\cite{Pellegrino2014a,Jenkins_thermshift,Jennewein_trans,Jenkins_long16}. If the sublevel $M_j$ of the electronic ground state $F=2$ is occupied, the optical
response of the atoms in that sublevel is described by three amplitudes $\Pc_{M_j,M_j+\sigma}^{(j)}$. Here the light polarization leads to three components $\sigma = -1,0,1$ and
$\Pc_{M_j,\eta}^{(j)}$ is the atomic polarization density amplitude for the $\ket{g,M_j}_j \leftrightarrow \ket{e,\eta}_j$ transition~\cite{Jenkins_long16}.  We consider the limit of low light
intensity and the electric dipole moment of the $J=0\rightarrow J'=1$ case is then replaced by the multilevel version
\begin{equation}
  \label{eq:dip_with_degeneracy}
  \spvec{d}_j(t) = \Dc \sum_{\sigma=-1}^{1}
  \cgshort{M_j}{\eta}{\sigma} \pola_{\sigma}\Pc_{M_j,\eta}^{(j)}(t)  \,\textrm{,}
\end{equation}
where $\cgshort{M}{\eta}{\sigma} \equiv \langle
F_e\eta;1F_g|F_gM;1\sigma\rangle$ are Clebsch-Gordan coefficients and the angular momenta of the ground and excited states are denoted by $F_g$ and $F_e$, respectively.

As a result, the equation for the polarization density amplitudes [Eq.~(1) in the main text] is replaced by
\begin{eqnarray}
  \label{eq:3}
  \frac{d}{dt} \Pc_{M_j\eta}^{(j)}& = &\left(i\Delta_{M_j\eta}
  - \gamma\right) \Pc_{M_j\eta}^{(j)} \nonumber\\
                                         & &   +i \frac{\xi}{{\cal D}}
                                             \cgshort{M_j}{\eta}{\sigma}
                                             \pola_{\sigma}^* \cdot \eo
                                             {\bf E}_{\rm ext}(\rv_j) \text{.}
\end{eqnarray}
In the presence of a magnetic field,  the Zeeman shifts of the different levels can be included in the detuning as in
\begin{equation}
\Delta_{M\eta} \equiv \omega - \big[\omega_0 +
\frac{\mu_BB}{\hbar} (g_eM - g_g\nu)\big]\,,
\end{equation}
where $\omega_0$ is the resonance frequency unperturbed by the magnetic bias
field.

Despite the multilevel structure, the coupled dynamics of Eq.~\eqref{eq:3} leads to the similar effective two-mode model as in the case of the $J=0\rightarrow J'=1$ transition, except that the transitions are weighted by the (possibly) different Clebsch-Gordan coefficients. Consequently, the subradiant state can be excited even in a multilevel case. However, it may be more challenging to set the Zeeman shifts $\bar\delta=0 $ after the dynamics has reached the steady state with a significant subradiant state excitation and when the incident field is turned off. If  $\bar\delta\neq 0 $, the collective excitation $ \Pc_P$ is still coupled to $ \Pc_I$ according to Eq.~(4a) in the main text, and can decay more rapidly via $ \Pc_I$. Achieving $\bar\delta=0$ may be simpler by using AC Stark shifts or if only the $m_F=0$ sublevel is occupied, such that only the electronically excited states $m_F=-1,0,+1$ are driven.

The simulations that we have presented are exact for stationary atoms for the  $J=0\rightarrow J'=1$ transition. For the multilevel case the simulation description includes the dipole-dipole interactions classically but ignores quantum fluctuations between the ground levels~\cite{Lee16}. These can be incorporated to the lowest order as an internal-state correlation functions~\cite{Lee16}. The similar techniques can also be applied when the saturation of the electronically excited levels is no longer negligible and the response of the atoms becomes nonlinear.

\end{document}